# Field-free magnetization reversal by spin-Hall effect and exchange bias

A. van den Brink[*,1], G. Vermijs,[1] A. Solignac,[1,2] J. Koo,[1] J.T. Kohlhepp,[1] H.J.M. Swagten[1], and B. Koopmans[1]

[1] Eindhoven University of Technology, PO Box 513, 5600 MB Eindhoven, The Netherlands

[2] DSM/IRAMIS/SPEC, CNRS UMR 3680, CEA Saclay, 91191 Gif sur Yvette cedex, France

**Magnetic random-access memory (MRAM) driven by spin-transfer torque (STT)[1,2] is a major contender for future memory applications[3,4,5]. The energy dissipation involved in writing remains problematic[6], even with the advent of more efficient perpendicular magnetic anisotropy (PMA) devices[7]. A promising alternative switching mechanism employs spin-orbit torques[8] and the spin-Hall effect[9,10] (SHE) in particular, but additional symmetry breaking is required to achieve deterministic switching in PMA devices. Currently used methods rely on in-plane magnetic fields[11,12] or anisotropy gradients[13], which are not suitable for practical applications. Here, we interface the magnetic layer with an anti-ferromagnetic material. An in-plane exchange bias (EB) is created, and shown to enable field-free SHE-driven magnetization reversal of a perpendicularly magnetized Pt/Co/IrMn structure. Aside from the potential technological implications, our experiment provides additional insight into the local spin structure at the ferromagnetic/anti-ferromagnetic interface.**

Research efforts to improve upon the STT writing paradigm explore the use of electric fields[14], multi-ferroics[15], perpendicular polarizers[16], and spin-orbit torques[8]. The latter category is dominated by devices employing the spin-Hall effect[9,10,17], which has been shown to be a viable method of spin injection in recent experiments[11,12,18,19]. Magnetization reversal using only SHE was demonstrated for in-plane magnetized magnetic tunnel junctions (MTJs)[18], but remains challenging in perpendicularly magnetized MTJs, which are more relevant due to higher data storage density. Additional symmetry-breaking is required to allow the in-plane polarized spin current generated from the SHE to deterministically switch out-of-plane magnetized devices. We address this issue by interfacing the perpendicularly magnetized layer with an anti-ferromagnetic material, creating an in-plane EB along the current flow direction. We demonstrate SHE-driven magnetization reversal using only the intrinsic in-plane magnetic field caused by this EB.

---

[*] Author to whom correspondence should be addressed. Electronic mail: a.v.d.brink@tue.nl

Experiments were performed on Ta (1) / Pt (3) / Co (0.7) / Pt (0.3) / IrMn (6) / TaOx (1.5) stacks (nominal thicknesses in nm), patterned into Hall crosses. Samples were field-cooled to set the EB along the $+\hat{y}$ direction (see Figure 1), as described in the Methods section. The presence of both PMA and in-plane EB was verified by carrying out Magneto-Optic Kerr Effect (MOKE) and SQUID Vibrating Sample Magnetometry (SQUID-VSM) measurements on unstructured samples after annealing (see Supplementary Information). Out-of-plane MOKE measurements show square loops with $\mu_0 H_C \approx 40$ mT and negligible EB. In-plane SQUID-VSM measurement shows an EB of $\mu_0 H_{EB} \approx 50$ mT. Furthermore, the saturation magnetization is measured at $M_S = 1.2$ MA/m with $\mu_0 H_K \approx 1.0$ T, indicating a substantial PMA of $K_{eff} \approx 8.3 \cdot 10^5$ J/m$^3$.

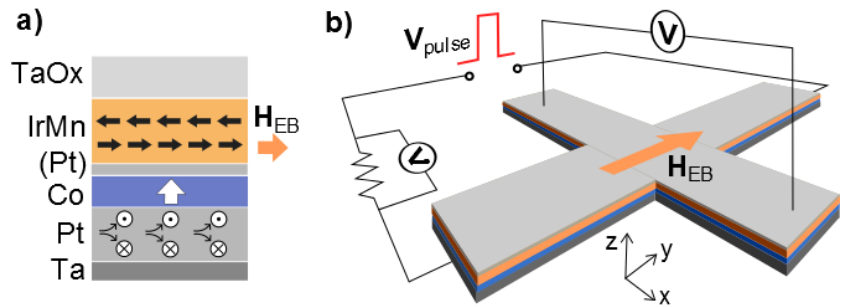

Figure 1: Schematic sample layout. a) Cross section of the deposited stack, showing the magnetic easy axis of the Co (white arrow), simplified spin structure of the IrMn (thick black arrows), exchange bias field (orange arrow), and spin current generated from a charge current running through the Pt (circles). b) Hall cross structure consisting of two 10 x 1 µm rectangles, and measurement scheme.

As a proof-of-principle, these samples are subjected to a sequence of current pulses along the $\hat{y}$ direction, in the absence of applied magnetic fields. Through the SHE, a current in the $\pm\hat{y}$ direction should generate a spin current polarized in the $\pm\hat{x}$ direction for positive spin-Hall angles, as in Pt[11]. Such a spin-current can switch the magnetization from $\pm\hat{z}$ to $\mp\hat{z}$, provided that both current density and the effective magnetic field along the $+\hat{y}$ direction are large enough. Switching in the other direction should occur only if the current polarity is reversed. We successfully demonstrate this behavior in our devices, using 50 μs current pulses ($J = 8 \cdot 10^{11}$ A/m$^2$) in the sequence shown in Figure 2a. No external magnetic field is present during this measurement. Deterministic switching is clearly observed upon reversing the current polarity, as seen in both anomalous Hall effect resistance ($R_{AHE}$) and MOKE measurements (Figure 2). Moreover, subsequent pulses of equal polarity have little effect on the magnetization. Furthermore, varying the pulse duration between 1 – 100 μs was found not to affect the end result significantly. From this proof-of-principle measurement, it is evident that the EB provides sufficient effective magnetic field to facilitate deterministic SHE-driven magnetization reversal.

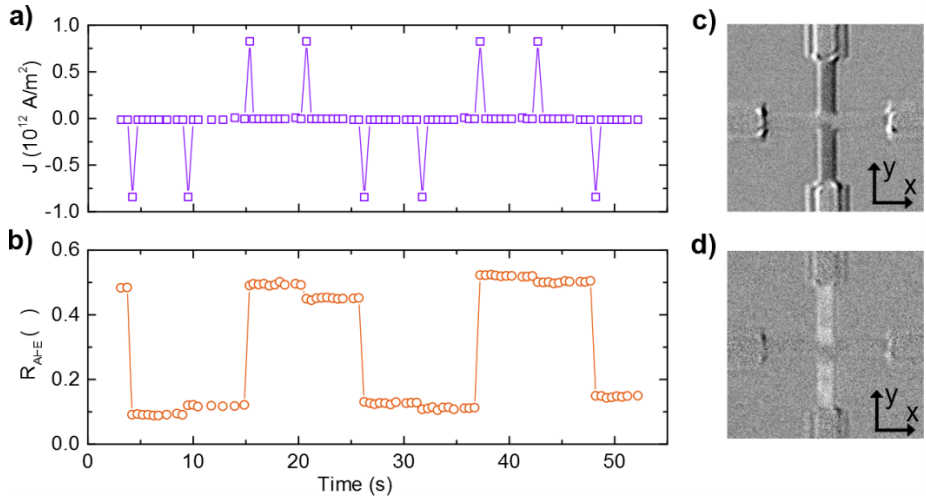

**Figure 2: Switching by current pulses. a)** Applied current density pulse pattern, and **b)** resulting anomalous Hall resistance. Switching is observed for one current polarity in either state, without any applied magnetic field. Differential Kerr microscopy images of the microwire after switching to **c)** the low $R_{AHE}$ state and **d)** the high $R_{AHE}$ state confirm the magnetization reversal.

Two more subtle features, visible in Figure 2, were found to be reproducible and require further investigation. First, the magnetization shows a small jump after repeated current pulses of the same polarity, which is unexpected. Second, the MOKE images suggest that magnetization reversal in the center of the Hall cross is less complete than outside this region. Taking into account that the current density is ~30% lower in the center of the Hall cross (see Supplementary Information), it appears that magnetization reversal in the absence of magnetic fields is incomplete, especially at lower current densities.

To explore this effect in more detail, we sweep the pulse current density from high negative values to high positive values and back. Additionally, we apply a magnetic field $B_y$ along the $\hat{y}$ direction to investigate how this affects the magnetization reversal. The resulting $R_{AHE}(J_{pulse})$ curves (Figure 3) show several interesting features.

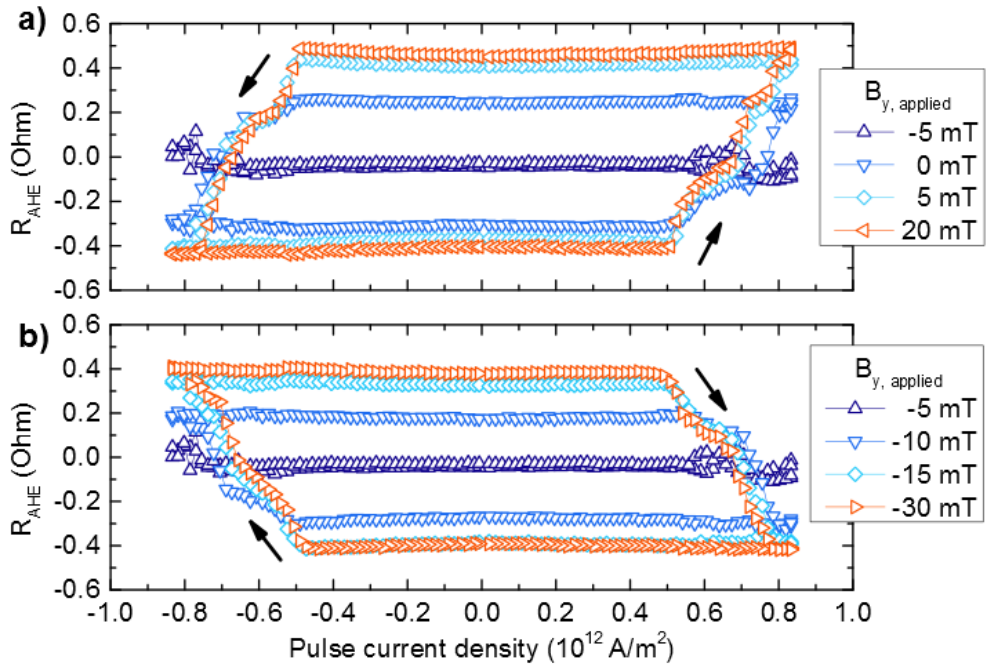

**Figure 3: Current density sweeps.** Anomalous Hall resistance $R_{AHE}$ measured during pulse current density sweeps for various applied in-plane magnetic fields. The magnetic field enhances deterministic **a)** upward and **b)** downward switching. The arrows indicate the sweep direction.

The total change in magnetization after a current density sweep, $\Delta R_{AHE}$, is found to strongly depend on $B_y$. For $B_y$ = -5 mT, we find that $\Delta R_{AHE}$ is negligible, implying a complete absence of deterministic switching. This result is expected for a spin-Hall current in the absence of an effective magnetic field, suggesting that the effective EB field is compensated by $B_y$ at this point. Note that this compensation point is not equal to the exchange bias field, as will be discussed later. Increasing $B_y$ in either direction is seen to gradually increase $\Delta R_{AHE}$: partial reversal is observed in the range -15 – 5 mT. This behavior is markedly different from devices without an EB, which have been shown to switch abruptly at a certain critical field[11,12].

Furthermore, a finite slope is clearly observed in the switching loops, representing a gradual change in $R_{AHE}$ for increasing $J_{pulse}$. This suggests that the magnetization reversal is not uniform but occurs in many small domains, each with a different critical current density for deterministic switching. Again, this behavior is radically different from samples without an EB, which show more sudden magnetization reversal (see Supplementary Information).

Finally, the current density required for magnetization reversal is identical for up-down and down-up switching, confirming that there is no preferential direction along the $\hat{z}$ axis. The vertical offset is negligible in all loops, indicating that the entire measured region is affected by the current. For the $B_y$ = -5 mT trace, for instance, this implies that a large current density produces equal amounts of up- and down magnetized domains, such that $R_{AHE} = 0$.

To further explore magnetization reversal driven by SHE and in-plane EB, we systematically measure vary the pulse current density and assisting magnetic field, both parallel and perpendicular to the EB direction. For each combination of field and pulse current density, the magnetization is first saturated in the $-\hat{z}$ direction. The change in $R_{AHE}$ before and after pulse application is measured and normalized to the largest recorded $\Delta R_{AHE}$, resulting in the phase diagrams shown in Figure 4, panels a and b. The diagrams agree with SHE-driven switching experiments[11,12] and provide several key insights into the effect of the EB, as detailed below.

First, we look at the $B_{IP} = 0$ traces in the phase diagrams. Confirming the proof-of-principle result, near-complete magnetization reversal is observed for strong current pulses along the exchange bias direction (Figure 4a). Furthermore, a maximum of 50% magnetization reversal is attained when measuring perpendicular to the EB direction (Figure 4b) even for high current densities, indicating random rather than deterministic switching.

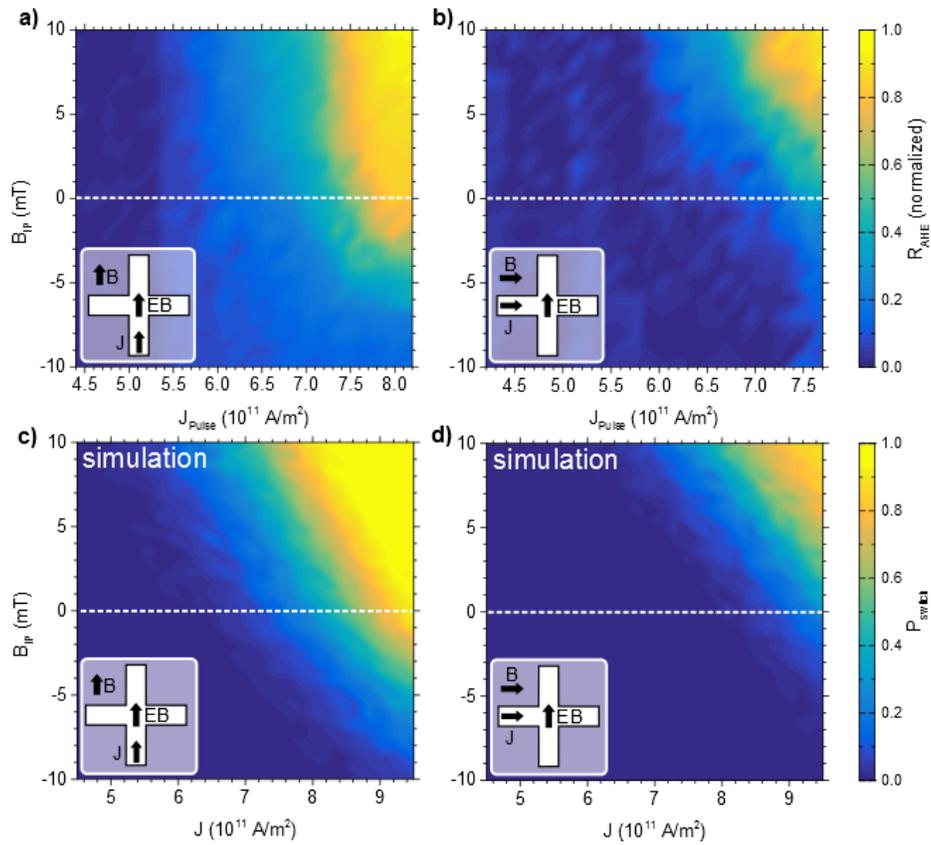

**Figure 4: Magnetization reversal as a function of current density *J* and collinear applied magnetic field $B_{IP}$.** Measurements show the normalized change in $R_{AHE}$ with *J* and $B_{IP}$ **a)** parallel and **b)** perpendicular to the exchange bias direction. The small difference in horizontal axes is caused by a resistance difference between the two directions of the Hall cross. Evaluation of the LLG equation, including local variations in the EB direction and magnitude, produces comparable phase diagrams for both the **c)** parallel and **d)** perpendicular configurations.

Second, the perpendicular-to-EB measurement resembles the parallel-to-EB measurement shifted vertically by $B_{IP} \approx 6$ mT; close to the effective EB observed in Figure 3. However, for intermediate current densities $\Delta R_{AHE}$ is larger parallel to the EB, as can be seen from the light blue area in Figure 4a. This implies that partial magnetization reversal, at intermediate current densities, is also easier along the EB direction.

Third, we find that the phase diagrams can be reproduced by numerical evaluation of the Landau-Lifshitz-Gilbert (LLG) equation (see Supplementary Information) implementing the SHE as an in-plane polarized spin current and the EB as an effective magnetic field (Figure 4, panels c and d). Importantly, the agreement between simulations and experiments is improved by selecting the EB magnitude and direction from appropriate distributions, as discussed below.

Concluding our measurements, deterministic switching of perpendicular magnetization by an in-plane current was demonstrated in the absence of magnetic fields. The magnetization reversal process is not complete, however, as concluded from measurements using an additional in-plane magnetic field. Partial switching appears to be intrinsic to SHE-driven magnetization reversal under small applied magnetic field. We believe that the physical origin of this effect must be sought in the local structure of the anti-ferromagnetic layer, which produces conditions subtly different from an applied magnetic field, which is inherently homogeneous. Sputtered IrMn has a poly-crystalline morphology[20] which complicates the simplistic picture of exchange bias painted in Figure 1a. During annealing, anti-ferromagnetic spins align to the field-cooling direction on average, but the actual spin direction within a grain is bound to local crystallographic axes[20] as sketched in Figure 5a. Furthermore, variations in grain size and orientation affect the local magnitude of the exchange bias[21]. This local spin structure, present in any exchange biased system, appears to affect SHE-driven magnetization reversal especially.

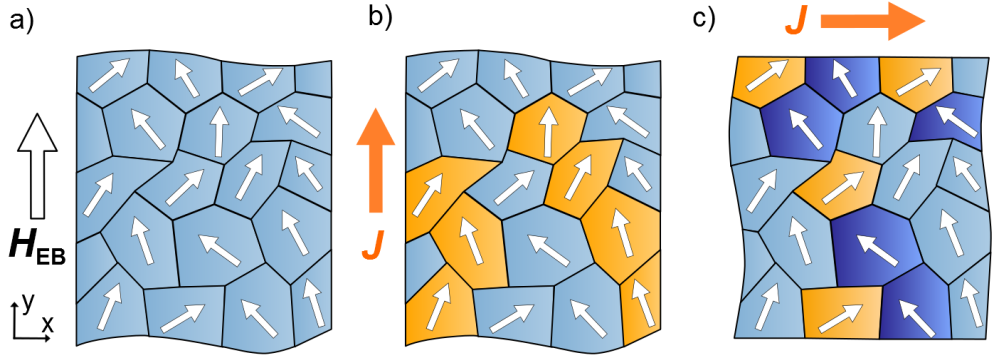

**Figure 5: Simplified two-dimensional sketches of grains within the anti-ferromagnetic layer.** The small arrows represent the uncompensated spin direction in each grain at the interface with the ferromagnet. **a)** Situation after field-cooling. Partial magnetization reversal occurs in the adjacent ferromagnetic layer after **b)** a current pulse along the EB direction or **c)** a current pulse perpendicular to the EB direction. Switched regions are indicated in orange, blocked regions in dark blue.

A current pulse can induce deterministic switching via SHE only if there is sufficient effective magnetic field along the current direction. We propose that, at a given current density and small applied in-plane field, these conditions hold only for a subset of regions where the local uncompensated spin direction has sufficient component along the current direction, as illustrated in Figure 5b. This explains why partial magnetization reversal is observed at small in-plane magnetic fields. Furthermore, grains can exist where the local exchange bias is against the current flow direction if one measures perpendicular to the EB direction (Figure 5c). Magnetization reversal is suppressed in such grains, which explains the reduced $\Delta R_{AHE}$ observed in Figure 4b for intermediate current densities.

As mentioned before, our experiments can be reproduced by numerical evaluation of the LLG equation. We implement the local spin structure of the anti-ferromagnetic layer by averaging over many simulations while drawing the EB direction from a distribution appropriate for a cubic poly-crystalline material. This produces a range of applied fields and current densities where partial magnetization reversal occurs, significantly improving the agreement with experiments over simulations with a uniform exchange bias of 5 mT. The agreement is further improved by drawing the local EB magnitude from a $\chi_3$-distribution to account for grain size variations (see Supplementary Information). SHE-driven magnetization reversal, aside from its technological relevance, may thus provides a unique tool in understanding the local spin structure at ferromagnetic/anti-ferromagnetic interfaces.

Finally, the apparent distribution in EB magnitude and direction partially explains the discrepancy between the EB field of 50 mT observed in SQUID-VSM measurements and the 5 mT effective in-plane field observed in current-driven switching experiments. In addition, it is known that patterned structures can exhibit reduced EB[22], and the used lift-off process may reduce the film quality. Improving fabrication conditions to obtain a more uniform (ideally single-crystalline) anti-ferromagnetic layer should resolve this issue, allowing for reliable field-free binary switching using the SHE. Future memory devices may even employ the SHE of anti-ferromagnetic metals[23] to enable read-out of the magnetic state using an MTJ structure.

In summary, we have demonstrated field-free SHE-driven magnetization reversal by interfacing an out-of-plane magnetized Co layer with a Pt spin-Hall injection layer and an IrMn exchange-biasing layer. A proof-of-principle measurement shows field-free switching and exhibits all expected symmetries. The amount of magnetization reversal is found to increase when applying an additional in-plane magnetic field. This observation can be attributed to the poly-crystalline nature of the anti-ferromagnet, as confirmed by simulations. Improving the crystalline structure of the anti-ferromagnetic layer could lead to reliable binary switching. We believe that these measurements provide a significant breakthrough in applied spintronics, as well as a unique probe for the local spin structure of poly-crystalline anti-ferromagnetic materials.

**METHODS**

**Sample fabrication.** Samples were fabricated on polished, thermally oxidized silicon substrates using DC sputtering at a base pressure around $10^{-8}$ mbar, The deposited stack (Figure 1a) consists of Ta (1) / Pt (3) / Co (0.7) / Pt (0.3) / $Ir_{20}Mn_{80}$ (6) / TaOx (1.5), with nominal thicknesses in nm. The Pt dusting layer was inserted to enhance the PMA and was found not to be detrimental to the EB, in agreement with the literature[24]. Layer thicknesses were chosen after careful optimization, as detailed in the Supplementary Information. Using a lift-off electron-beam lithography procedure, the stack is patterned into Hall crosses (Figure 1b) consisting of two overlapping 10 x 1 μm rectangles. A small pad at each extremity of the Hall cross connects to thick Ti/Au electrodes (not shown in the figure) to allow for electrical contact. The completed structures are then placed in a 2.0 T in-plane magnetic field along one of the Hall bar axes, annealed at 225 °C for 30 minutes, and finally field-cooled to set the EB direction.

**Experimental set-up.** The magnetization reversal process was studied using a Kerr microscope in polar mode, allowing for high-resolution digital imaging of the out-of-plane magnetization component. Additionally, an Agilent 33250A pulse generator was used to apply voltage pulses and a small DC voltage to allow for $R_{AHE}$ measurements, providing an accurate measure of the average out-of-plane magnetization in the junction area. The pulse current could be determined by monitoring the voltage drop over a resistor in series with the device. Current densities are computed by dividing the current over the total metallic cross-sectional area of the microwire, which is 11 nm x 1 μm.

# REFERENCES

[1] Slonczewski, J. C. Current-driven excitation of magnetic multilayers. *J. Magn. Magn. Mater.* **159,** L1–L7 (1996).

[2] Berger, L. Emission of spin waves by a magnetic multilayer traversed by a current. *Phys. Rev. B* **54,** 9353 (1996).

[3] Wolf, S. A., Lu, J., Stan, M. R., Chen, E. & Treger, D. M. The Promise of Nanomagnetics and Spintronics for Future Logic and Universal Memory. *Proc. IEEE* **98,** 2155–2168 (2010).

[4] Apalkov, D. *et al.* Spin-transfer torque magnetic random access memory (STT-MRAM). *ACM J. Emerg. Technol. Comput. Syst.* **9,** 1–35 (2013).

[5] Wang, K. L., Alzate, J. G. & Khalili Amiri, P. Low-power non-volatile spintronic memory: STT-RAM and beyond. *J. Phys. D. Appl. Phys.* **46,** 074003 (2013).

[6] Chen, E. *et al.* Advances and Future Prospects of Spin-Transfer Torque Random Access Memory. *IEEE Trans. Magn.* **46,** 1873–1878 (2010).

[7] Ikeda, S. *et al.* A perpendicular-anisotropy CoFeB-MgO magnetic tunnel junction. *Nat. Mater.* **9,** 721–4 (2010).

[8] Brataas, A. & Hals, K. M. D. Spin-orbit torques in action. *Nat. Nanotechnol.* **9,** 86–8 (2014).

[9] D'yakonov, M. I. & V.I. Perel', V. I. Possibility of Orienting Electron Spins with Current. *Sov. Phys. JETP Lett.* **13**, 467 (1971).

[10] Hirsch, J. Spin Hall Effect. *Phys. Rev. Lett.* **83,** 1834–1837 (1999).

[11] Liu, L., Lee, O., Gudmundsen, T., Ralph, D. & Buhrman, R. A. Current-Induced Switching of Perpendicularly Magnetized Magnetic Layers Using Spin Torque from the Spin Hall Effect. *Phys. Rev. Lett.* **109,** 1–5 (2012).

[12] Miron, I. M. *et al.* Perpendicular switching of a single ferromagnetic layer induced by in-plane current injection. *Nature* **476,** 189–93 (2011).

[13] Akyol, M. *et al.* Current-induced spin-orbit torque switching of perpendicularly magnetized Hf|CoFeB|MgO and Hf|CoFeB|TaOx structures. *Appl. Phys. Lett.* **106,** 162409 (2015).

[14] Matsukura, F., Tokura, Y. & Ohno, H. Control of magnetism by electric fields. *Nat. Nanotechnol.* **10,** 209–220 (2015).

[15] Heron, J. T. *et al.* Deterministic switching of ferromagnetism at room temperature using an electric field. *Nature* **516,** 370–373 (2014).

[16] Liu, H. Spin-transfer magnetic random access memory devices with an orthogonal polarizing layer. *Mod. Phys. Lett. B* **28**, 1430005 (2014).

[17] Van den Brink, A. *et al.* Spin-Hall-assisted magnetic random access memory. *Appl. Phys. Lett.* **104,** (2014).

[18] Liu, L. *et al.* Spin-Torque Switching with the Giant Spin Hall Effect of Tantalum. *Science* **336,** 555–558 (2012).

[19] Haazen, P. P. J. *et al.* Domain wall depinning governed by the spin Hall effect. *Nat. Mater.* **12,** 299–303 (2013).

[20] O'Grady, K., Fernandez-Outon, L. E. & Vallejo-Fernandez, G. A new paradigm for exchange bias in polycrystalline thin films. *J. Magn. Magn. Mater.* **322,** 883–899 (2010).

[21] Scholl, A. *et al.* Domain-size-dependent exchange bias in Co/LaFeO$_3$. *Appl. Phys. Lett.* **85,** 4085 (2004).

[22] Nogués, J. & Schuller, I. K. Exchange bias. *J. Magn. Magn. Mater.* **192,** 203–232 (1999).

[23] Zhang, W. *et al.* Spin Hall Effects in Metallic Antiferromagnets. *Phys. Rev. Lett.* **113,** 196602–196608 (2014).

[24] Van Dijken, S., Besnier, M., Moritz, J. & Coey, J. M. D. IrMn as exchange-biasing material in systems with perpendicular magnetic anisotropy. *J. Appl. Phys.* **97,** 10–12 (2005).

# Field-free magnetization reversal by spin-Hall effect and exchange bias

## Supplementary Information

A. van den Brink, G. Vermijs, A. Solignac, J. Koo, J.T. Kohlhepp, H.J.M. Swagten, and B. Koopmans

## I. THIN FILM MAGNETIZATION MEASUREMENTS

Thin film magnetic characteristics were studied using polar Magneto-Optic Kerr Effect (MOKE) and SQUID Vibrating Sample Magnetometry (SQUID-VSM) measurements, both before and after the field-cooling process. Here, we show the data obtained for the Ta (1) / Pt (3) / Co (0.7) / Pt (0.3) / IrMn(6) / TaOx(1.5) stack (nominal thicknesses in nm) discussed in the main text.

After deposition, MOKE measurements show a typical double-loop behavior (Figure S6a). A random pattern of up- and downward out-of-plane (OOP) magnetization arises in the Co layer during deposition, and is transferred to the IrMn layer. This creates regions of positive and negative exchange bias (EB) as seen in the MOKE loop. A uniform OOP EB can be obtained by heating the sample to 225 °C while applying an OOP magnetic field larger than the coercive field. This is shown in Figure S6b; a MOKE measurement taken after OOP field cooling at 0.2 T.

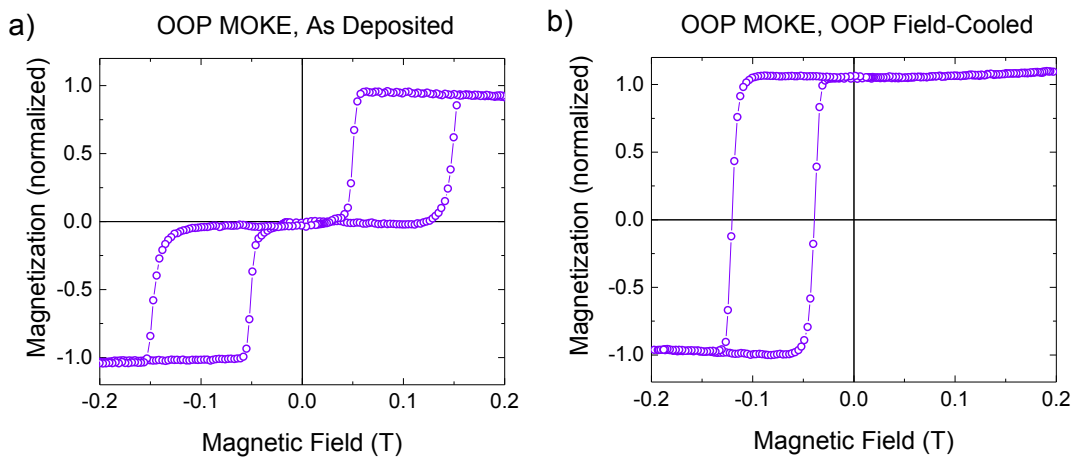

Figure S6: Thin film magnetization versus magnetic field measured by polar MOKE along the out-of-plane axis in the a) as-deposited and b) out-of-plane field-cooled states.

Creating an in-plane EB in samples with perpendicular magnetic anisotropy (PMA) is more difficult, as the intrinsic anisotropy field needs to be overcome to force the magnetization of the Co layer in the in-plane direction during field-cooling. We therefore applied a large in-plane field of 2.0 T while heating the sample to 225 °C and field-cooling over a period of 30 minutes. Afterwards, OOP MOKE measurements (Figure S7a) show full remanence, a substantial coercivity $\mu_0 H_c \approx 40$ mT, and negligible EB in the out-of-plane direction. The squareness of the loop is clear evidence for a substantial perpendicular magnetic anisotropy. In-plane SQUID-VSM measurements (Figure S7b) show an in-plane EB of $\mu_0 H_{EB} \approx 50$ mT and a saturation magnetization $M_s = 1.2$ MA/m with $\mu_0 H_K \approx 1$ T, indicating an effective PMA of $K_{eff} \approx 8.3 \cdot 10^5$ J/m$^3$. A slight opening is visible in the SQUID-VSM cycle, probably caused by a slight misalignment between the sample surface and the measurement direction.

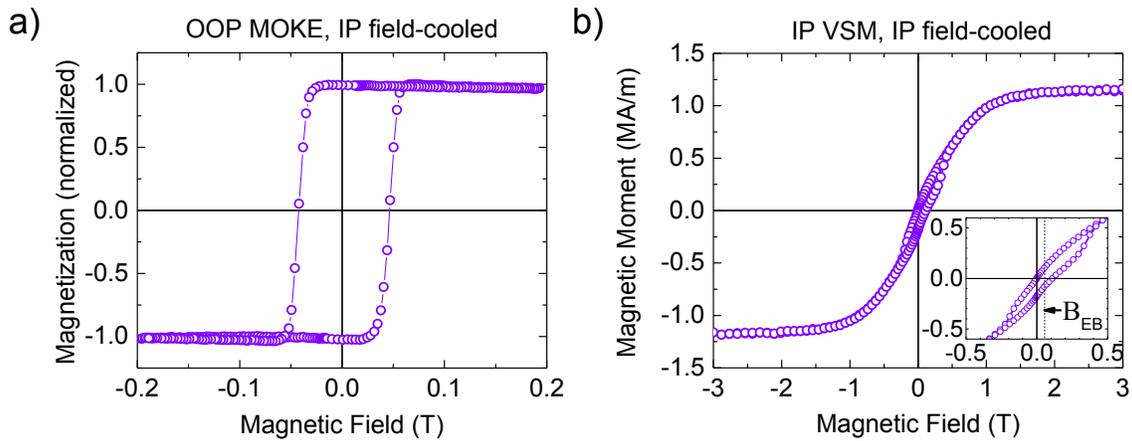

Figure S7: Thin film magnetization versus magnetic field measured by a) polar MOKE along the out-of-plane axis and b) SQUID-VSM along the in-plane direction. The sample shows fully remanent out-of-plane magnetization with a coercive field of 40 mT as well as an exchange bias field of 50 mT in the in-plane direction (see inset).

## II. STACK OPTIMIZATION

The composition of the material stack used in our experiments is the result of careful optimization of all layer thicknesses. Relevant results are presented here, to provide an overview of the effect of each layer thickness.

Deposition starts with a Ta seed layer, which is commonly used to improve film quality[25] and was found to significantly increase the PMA in our samples. The thickness of this buffer layer was minimized to reduce current shunting effects. This reduces the PMA, but we found that a 1 nm Ta seed layer suffices for our measurements.

The 3 nm Pt thickness should maximize the spin-Hall effect (SHE) efficiency; see section III.

The Co layer was chosen as thin as possible, to maximize both the PMA and the susceptibility to spin currents injected from the interface. MOKE measurements were performed on a sample with a variable Co thickness, which was subjected to the in-plane field cooling process. As shown in Figure S8, a thickness of 0.7 nm yields the largest coercivity and full remanence ($M(0)/M_s$), indicating that a substantial PMA is obtained for this Co thickness.

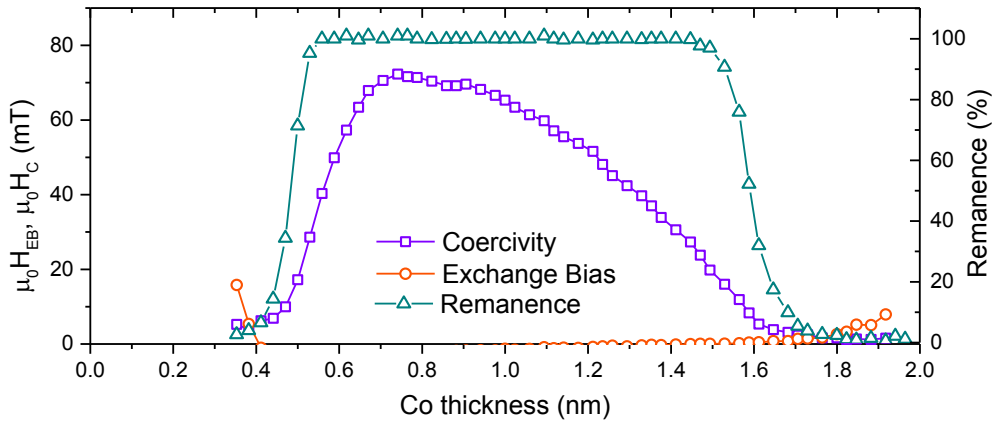

Figure S8: graphs of the exchange bias $H_{EB}$, coercive field $H_c$, and remanence as a function of cobalt thickness in a Ta (3) / Pt (3) / Co (0-2) / Pt (0.3) / IrMn (6) / Pt (3) wedge sample, measured using polar MOKE in the out-of-plane direction after annealing in the in-plane direction at 225°C in a 2T magnetic field for 30 minutes.

The thickness of the IrMn layer is crucial to obtain a large and stable EB. We created a sample with a variable IrMn thickness, and annealed it at 225°C in a 0.2 T out-of-plane magnetic field for 30 minutes. This allows us to measure the EB and coercivity as a function of IrMn thickness using polar MOKE, which we found to be a good measure for the properties of an in-plane annealed sample. As shown in Figure S9, the highest EB is obtained for an IrMn thickness of 6 nm. Note that the coercivity peak and negligible EB indicate that the EB is unstable for reduced thicknesses. The reduction in EB observed at higher thicknesses can probably be attributed to changes in microstructure or domain structure in the IrMn[26].

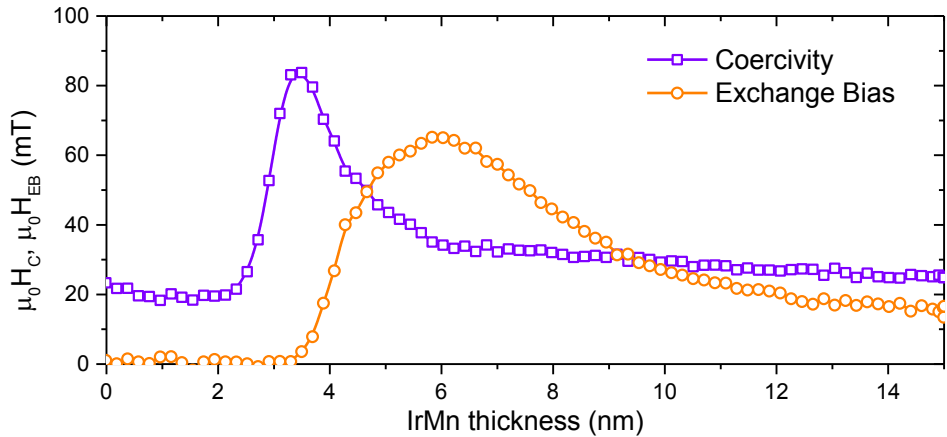

Figure S9: graphs of the exchange bias $H_{EB}$ and coercive field $H_c$ as a function of IrMn thickness in a Ta (3) / Pt (4) / Co (1.5) / IrMn (0-15) / Pt (2) wedge sample, measured along the out-of-plane direction after annealing in the out-of-plane direction at 225°C in a 0.2 T magnetic field for 30 minutes. The sample shows close to 100% remanence across the entire IrMn thickness range.

A 0.3 nm Pt dusting layer was inserted between the Co and IrMn layers to increase the PMA of the Co layer. Interestingly, this dusting layer was found to significantly reduce the chance of device breakdown at high current densities. Note that this thin layer is not expected to contribute significantly to the net spin current due to scattering effects, as discussed in section III.

Finally, the stack is capped with a 1.5 nm Ta layer which is allowed to oxidize naturally, producing a protective yet transparent and non-conductive capping layer.

## III. SPIN CURRENT CONSIDERATIONS

Owing to the spin-Hall effect, a vertical spin current density $J_s$ can be generated from a planar charge current $J_e$ in materials with a nonzero bulk spin-Hall angle $\theta_{SH} \equiv J_s/J_e$. For our thin Pt films, we use the reported[27] value of $\theta_{SH}= 0.07$. Note that extensive debate exists on this subject, which is beyond the scope of this publication. For ultrathin films, the thickness of the metallic layer affects the net spin current. Spin accumulations are created at the interfaces with adjacent layers, causing spin diffusion that reduces the net spin current significantly if the film thickness is of the order of the spin-diffusion length $\lambda_{sf}$. Following the approach of Liu et al.[27], we take $\lambda_{sf}$ = 1.4 nm for Pt and model the net spin current as

$$J_s = J_e \theta_{SH}(1 - \text{sech}\frac{d}{\lambda_{sf}}), \quad \text{(S1)}$$

where $d$ is the Pt layer thickness. From this perspective, a thicker Pt layer is beneficial as it improves the net spin current. However, this also increases the total electric current $I_e$ required to produce a certain current density $J_e$, which increases Joule heating and thus the risk of device breakdown. To solve this trade-off, we compute the spin current $J_s$ as a function of Pt thickness $d$ while constraining $J_e$ to maintain a constant total current $I_e$. The result of this computation is shown in Figure S10. Current shunting through other metallic layers in the stack is taken into account, using a basic calculation where the stack is regarded as a parallel resistor network with appropriate resistances based on bulk conductivities. The optimum value for the Pt thickness is thus determined to be between 3 and 4 nm.

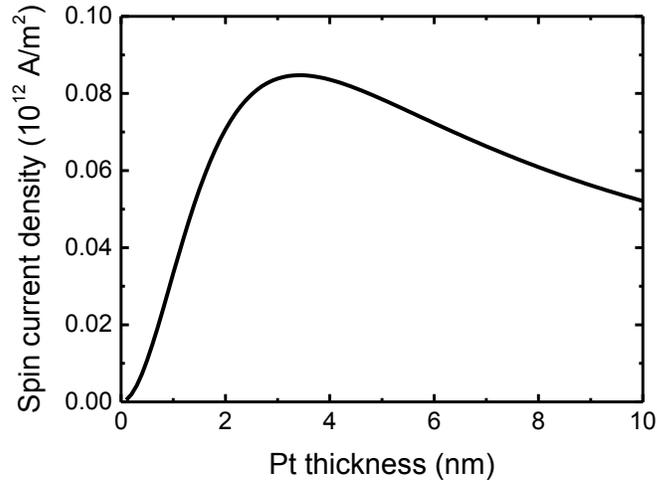

Figure S10: spin current density as a function of Pt thickness, computed for constant total current $I_e$. Spin diffusion and current shunting effects are taken into account using simple models. The optimum Pt thickness is found to be between 3 and 4 nm.

In our calculations, we only take into account the spin current generated from the thick Pt layer. This is justified, as contributions from the other layers are negligible. By the diffusion mechanism here, the contribution from the 0.3 nm Pt dusting layer is about thirty times smaller than the contribution from the 3 nm Pt layer. The Ta seed layer is also very thin, and its local conductivity will be significantly reduced due to elastic scattering at the substrate interface. Finally, although IrMn may exhibit a spin-Hall angle comparable to Pt[28], its conductivity is more than an order of magnitude lower. Therefore, its contribution to the total current density and total spin current is negligible compared to the Pt.

## IV. CURRENT SHUNTING EFFECT

As discussed in the main text, several experiments suggest that magnetization reversal in the center of the Hall cross occurs at a higher applied current than reversal in the rest of the microwire. To explain this behavior, we examine the current density distribution in the Hall cross using COMSOL simulations. Due to the current shunting through the inactive arms of the Hall cross, the current density in the center is reduced by roughly 30%, as shown in Figure S11. Referring to the main text, such a significant reduction is indeed expected to affect magnetization reversal in our experiments.

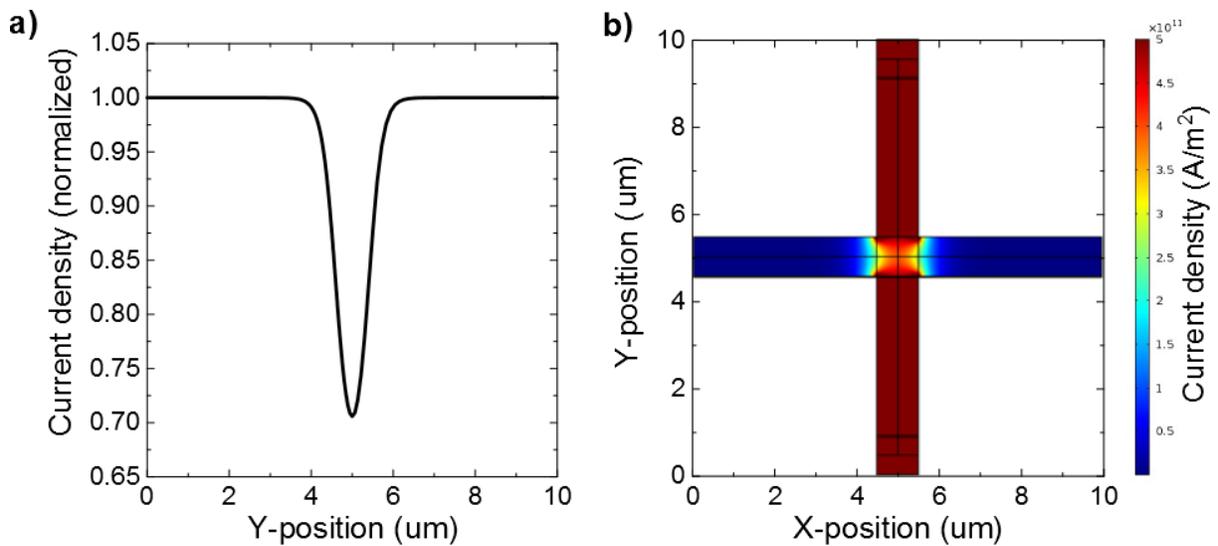

Figure S11: a) normalized current density along the central axis in the y-direction. The current density in the center is about 30% lower. b) 2D current density plot of the Hall cross, where current is flowing along the y-direction.

## V. PROOF-OF-PRINCIPLE WITHOUT DUSTING LAYER

The main text focuses on measurements performed in a Pt/Co/Pt/IrMn sample, where a Pt dusting layer was added between the Co and IrMn to improve the anisotropy. Similar measurements were performed on samples without dusting layers, with similar results. The proof-of-principle experiment performed on a Ta(0.5) / Pt (4) / Co (1.25) / IrMn (6) / TaOx (1.5) (nominal thicknesses in nm) is shown in Figure S12, and also exhibits field-free magnetization reversal driven by an in-plane current. These devices were not as robust against high current densities, so construction of a complete phase diagram proved impossible. The main text therefore focuses on a sample with a Pt dusting layer, which proved to be more stable at high current densities.

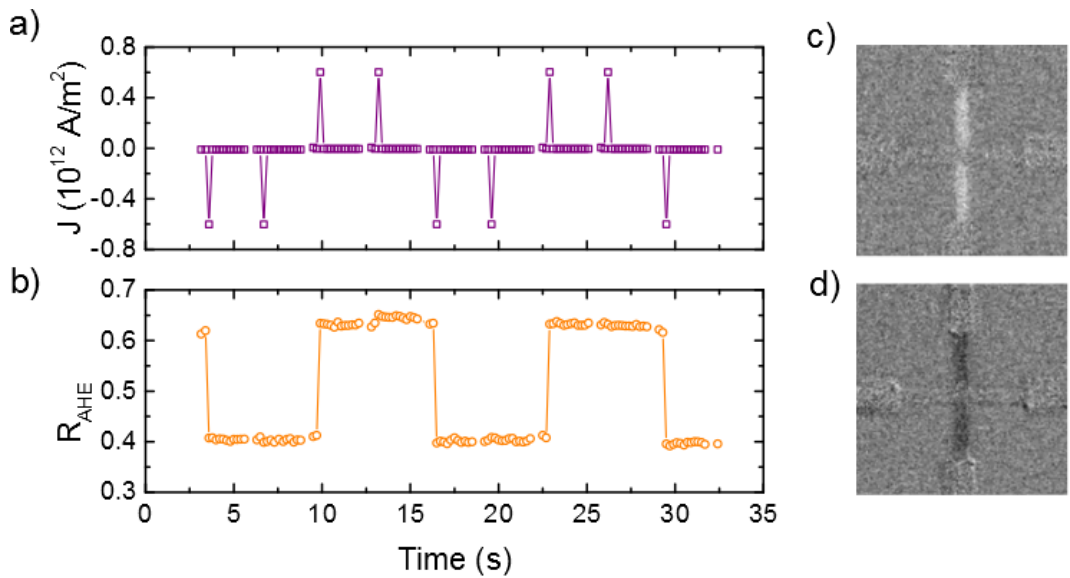

Figure S12: proof-of-principle measurement in a Ta (0.5) / Pt (4) / Co (1.25) / IrMn (6) / TaOx (1.5) sample, demonstrating that a) a train of current pulses produces b) magnetization switching as expected from the SHE. Kerr microscopy images show the magnetization after switching to c) the up-state and d) the down-state.

## VI. ON DOMAIN WALL PROPAGATION VERSUS DOMAIN NUCLEATION

As discussed in the main text, gradual magnetization reversal is observed when sweeping the current density from high negative to high positive values and back, even in the presence of additional in-plane magnetic fields. An explanation for this effect was offered, suggesting that the local spin structure of the IrMn causes a distribution of effective local magnetic fields, causing the observed behavior.

To test if the presence of IrMn indeed causes the observed gradual magnetization reversal, we created a Hall cross sample without an anti-ferromagnetic layer, composed of Ta (4) / Pt (3) / Co (1.2) / Ta(5) with nominal thicknesses in nm. In this case, the pulse current density sweep measurement produces abrupt switching, as shown in Figure S13. We conclude that the gradual magnetization reversal is not related to the device geometry, and must indeed by caused by the presence of IrMn in our samples.

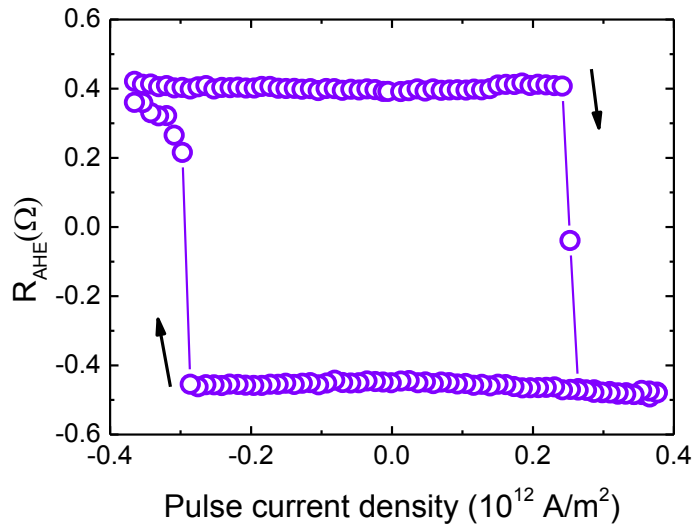

Figure S13: anomalous hall resistance $R_{AHE}$ of a Ta(4)/Pt(3)/Co(1.2)/Ta(5) microwire, recorded while sweeping the applied pulse current density from high negative values to high positive values and back under application of a -10 mT magnetic field along the current flow direction. Sudden magnetization reversal is observed at a critical current density below $3 \cdot 10^{11}$ A/m$^2$.

## VII. SIMULATION DETAILS

Following the approach of our earlier work[29], magnetization dynamics are simulated by solving the Landau–Lifshitz–Gilbert (LLG) equation[30]:

$$\frac{\partial \mathbf{M}}{\partial t} = -\gamma\mu_0(\mathbf{M} \times \mathbf{H}_{\text{eff}}) + \frac{\alpha}{M_s}\left(\mathbf{M} \times \frac{\partial \mathbf{M}}{\partial t}\right) + \frac{c_{\text{SHE}}}{M_s^2}(\mathbf{M} \times \hat{\boldsymbol{\sigma}}_{\text{SHE}} \times \mathbf{M}) \quad (S2)$$

with $\mathbf{M}$ the free layer magnetization, $\gamma$ the electron gyromagnetic ratio, $\mu_0$ the vacuum permeability, $\mathbf{H}_{\text{eff}}$ the effective magnetic field, $\alpha$ the Gilbert damping coefficient, and $M_s \equiv |\mathbf{M}|$ the saturation magnetization. The spin-Hall torque coefficient is given by $c_{\text{SHE}} = J_{\text{SHE}}\theta_{\text{SHE}}\hbar\gamma/(2ed)$, $J_{\text{SHE}}$ the spin-Hall effect current density running underneath the free layer, $\theta_{\text{SHE}}$ the spin-Hall angle, $\hbar$ the reduced Planck constant, $e$ the elementary charge, and $d$ the free magnetic layer thickness. The Oersted field generated by $J_{\text{SHE}}$ is approximated by that of an infinite surface current, whereas Joule heating and current shunting effects are neglected.

In simulating spin-Hall driven magnetization reversal in the presence of exchange bias, the effective field $\mathbf{H}_{\text{eff}}$ comprises four contributions: the applied magnetic field $\mathbf{H}_{\text{appl}}$, the exchange bias field $\mathbf{H}_{\text{EB}}$, the effective anisotropy field $\mathbf{H}_{\text{ani}} = 2K_U/(\mu_0 M_s)\hat{\mathbf{z}}$, with $K_U$ the uniaxial anisotropy energy density, and the demagnetizing field $\mathbf{H}_D$ which is approximated for an infinite thin film. Equation (S1) is solved numerically using an implicit midpoint rule scheme[31]. The SHE current is constant throughout each simulation, which last for 10 ns. The magnetization is considered 'switched' if the final magnetization vector $\mathbf{M}$ has a $\mathbf{z}$-component opposite in sign to the starting condition.

We set $K_U$ to $4.33 \cdot 10^5$ J/m$^3$ which is lower than the experimental value, to account for micromagnetic effects that produce lower effective switching fields. The used anisotropy yields a thermal stability of $\Delta \equiv K_{\text{eff}}V/(k_B T) = 60$ at room temperature for a 100x100 nm bit, with $K_{\text{eff}}$ the effective anisotropy after correcting for the demagnetization field. This shows that our effective anisotropy value is a reasonable assumption. Further notable parameters include $\alpha = 0.2$, $M_s = 1.0 \times 10^6$ Am$^{-1}$ for Co[32], $\theta_{SHE} = 0.07$ for Pt[27].

As mentioned in the main text, the local structure of the anti-ferromagnetic material can be approximated in simulations. Two modifications can be implemented, as discussed in the main text and elucidated below. The averaged effect of these local variations is computed by averaging over 256 simulations, each with different local exchange bias parameters.

First, the exchange bias direction can locally vary from the field-cooling direction. To simulate this, we draw the exchange bias direction from a uniform random distribution over the surface of a sphere. We then collapse this distribution between $-\pi/4 < \theta < \pi/4$ and $-\pi/4 < \psi < \pi/4$, where $\theta$ and $\psi$ are the azimuthal and elevation angle with respect to the $\hat{y}$ direction, respectively. The resulting distribution is constrained to 45° offset angles from the $\hat{y}$ direction, as illustrated in Figure S14. This simulates the field-cooling procedure, where the exchange bias direction is forced to the nearest cubic axis of each grain in the poly-crystalline material.

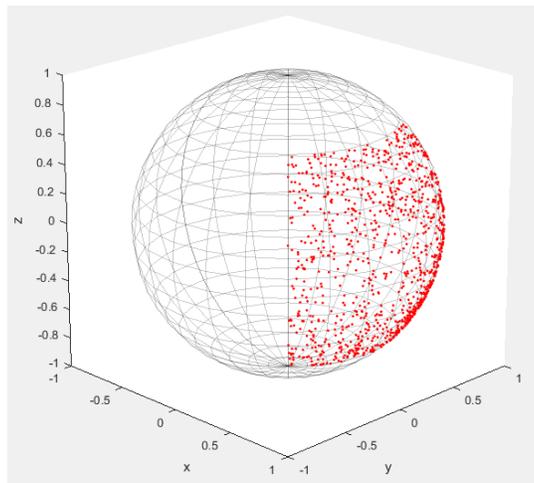

Figure S14: exchange bias direction distribution, plotted on the surface of a unit sphere

Second, we implement variations in the local exchange bias magnitude (due to grain size variations, for instance) by drawing it from a chi-distribution with 3 degrees of freedom ($\chi_3$, similar to the Maxwell-Boltzmann distribution) as plotted in Figure S15. The width of this distribution is chosen such that the simulated phase diagrams parallel and perpendicular to the exchange bias direction display a vertical shift similar to the one observed in experiments.

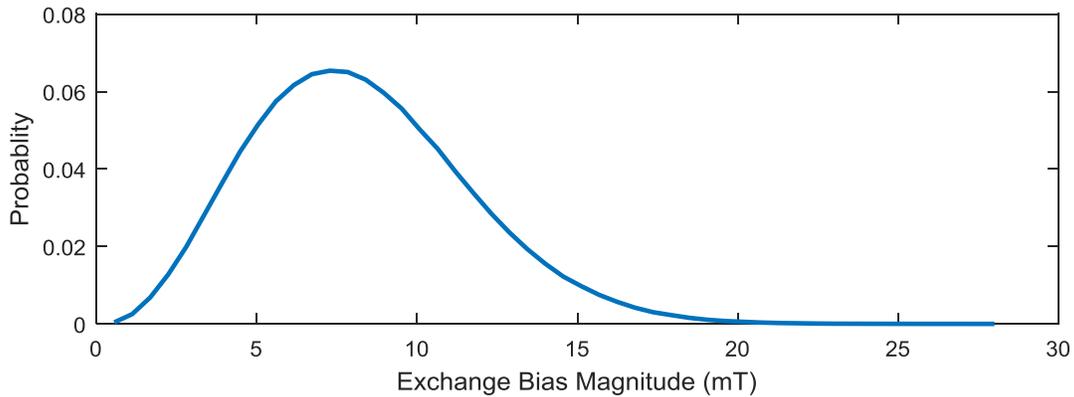

Figure S15: exchange bias magnitude distribution, based on a Maxwell-Boltzmann distribution.

Finally, note that our choice of distributions for the exchange bias direction and magnitude should be considered an Ansatz based upon our experimental data. Our model suffices to explain the important trends observed in our experiments (see section VIII), but the agreement is not perfect. This is particularly visible in the imperfect reproduction of the shape of the high-probability switching region along the exchange bias direction (c.f. Figure 4a and Figure 4c, main text). Further research may result in a more accurate description of the local exchange bias parameters.

## VIII. SIMULATION RESULTS

The simulations described in Section VII yield the phase diagrams shown in Figure S16. The left-hand (right-hand) panels show the situation where the current flow is along (perpendicular to) the mean EB direction. Compared to a uniform 5 mT exchange bias, both direction and magnitude variations improve the agreement with experimental data (c.f. Figure 4, main text).

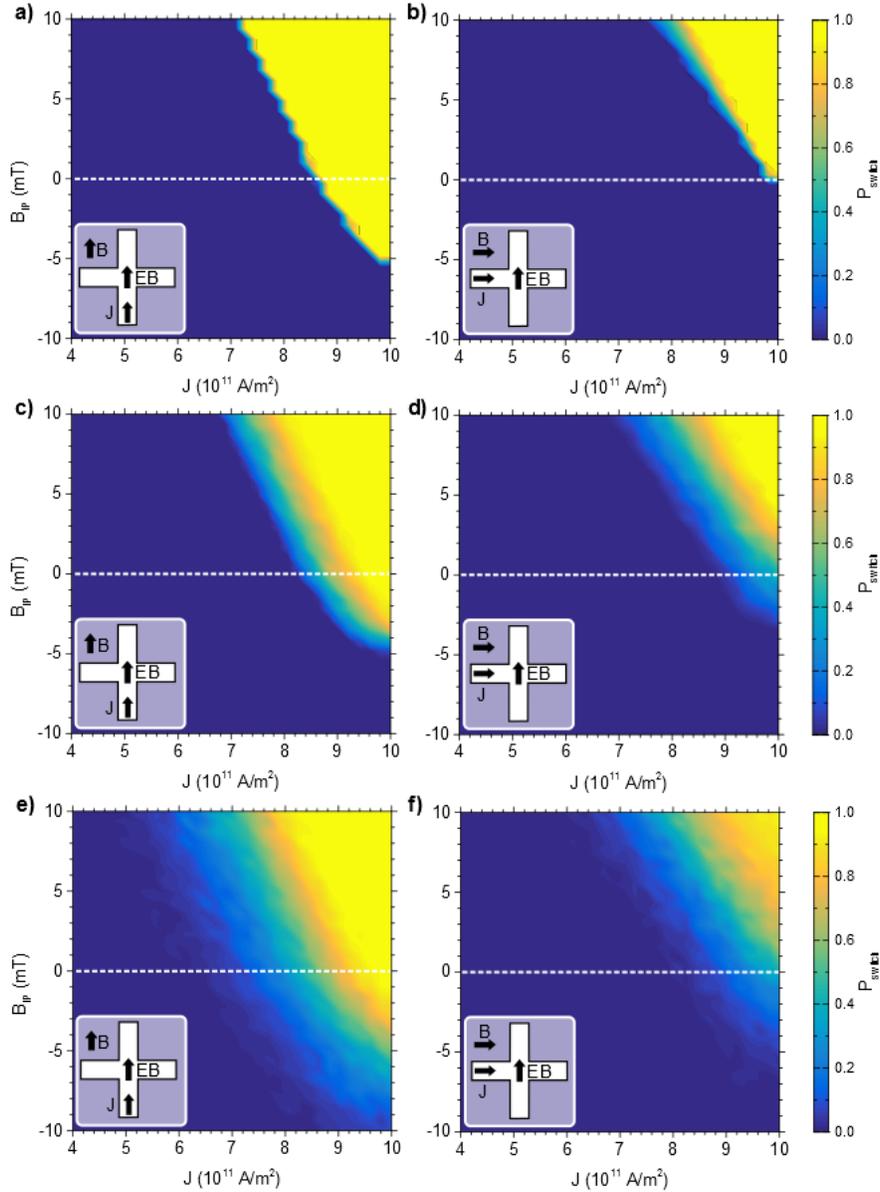

Figure S16: simulated switching probability $P_{switch}$ as a function of current density $J$ and magnetic field $B_{ip}$ along the current direction, in the presence of a,b) a uniform 5 mT EB, c,d) a 5 mT EB with angular spread, and e,f) a $\chi_3$-distributed EB with angular spread. For each $(J, B_{ip})$ point, $P_{switch}$ is computed by averaging over 256 simulations.

**REFERENCES (SUPPLEMENTARY INFORMATION)**

[25] Y.-T. Chen, Y.C. Lin, S.U. Jen, J.-Y. Tseng, and Y.D. Yao, J. Alloys Compd. **509** (18), 5587-5590 (2011).

[26] J. Nogués and I.K. Schuller, J. Magn. Magn. Mater. **192**, 203–232 (1999).

[27] L. Liu, O. Lee, T. Gudmundsen, D. C. Ralph, and R. A. Buhrman, Phys. Rev. Lett. **109** (9), 1–5 (2012).

[28] W. Zhang, M.B. Jungfleisch, W. Jiang, J.E. Pearson, and A. Hoffmann, Phys. Rev. Lett. **113**, 196602–196608 (2014).

[29] A. van den Brink, S. Cosemans, S. Cornelissen, M. Manfrini, A. Vaysset, W. Van Roy, T. Min, H.J.M. Swagten, and B. Koopmans, Appl. Phys. Lett. **104** 012403 (2014).

[30] R. Koch, J. Katine, and J. Sun, Phys. Rev. Lett. **92** (8), 088302 (2004).

[31] M. d'Aquino, C. Serpico, G. Coppola, I. D. Mayergoyz, and G. Bertotti, G., J. Appl. Phys. **99** (8), 08B905 (2006).

[32] L. Liu, O. Lee, T. Gudmundsen, D. C. Ralph, and R. A. Buhrman, Phys. Rev. Lett. **109** (9), 1–5 (2012).